\documentclass[final,5p,times,twocolumn]{elsarticle}

\usepackage{amsfonts,amssymb,amsmath,bm,empheq}
\usepackage[dvipsnames]{xcolor}
\usepackage{balance}

\usepackage[labelformat=simple]{subcaption}
\usepackage[colorlinks,citecolor=blue]{hyperref}
\usepackage{orcidlink}

\begin{document}

\begin{frontmatter}

\title{A fluid--peridynamic structure model of deformation and damage of microchannels}

\author[PUME]{Ziyu Wang\,\orcidlink{0000-0002-7749-3152}}
\ead{wang5600@purdue.edu}

\author[PUME]{Ivan C.\ Christov\corref{cor1}\,\orcidlink{0000-0001-8531-0531}}
\ead{christov@purdue.edu}
\ead[url]{https://tmnt-lab.org}
\cortext[cor1]{Corresponding author.}

\affiliation[PUME]{%
            organization={School of Mechanical Engineering, Purdue University},
            city={West Lafayette},
            postcode={47907}, 
            state={Indiana},
            country={USA}}

\begin{abstract}
Soft-walled microchannels arise in many applications, ranging from organ-on-a-chip platforms to soft-robotic actuators. However, despite extensive research on their static and dynamic response, the potential failure of these devices has not been addressed. To this end, we explore fluid--structure interaction in microchannels whose compliant top wall is governed by a nonlocal mechanical theory capable of simulating both deformation and material failure. We develop a one-dimensional model by coupling viscous flow under the lubrication approximation to a state-based peridynamic formulation of an Euler--Bernoulli beam. The peridynamic formulation enables the wall to be modeled as a genuinely nonlocal beam, and the integral form of its equation of motion remains valid whether the deformation field is smooth or contains discontinuities. Through the proposed computational model, we explore the steady and time-dependent behaviors of this fluid--peridynamic structure interaction. We rationalize the wave and damping dynamics observed in the simulations through a dispersion (linearized) analysis of the coupled system, finding that, with increasing nonlocal influence, wave propagation exhibits a clear departure from classical behavior, characterized by a gradual suppression of the phase velocity. The main contribution of our study is to outline the potential failure scenarios of the microchannel's soft wall under the hydrodynamic load of the flow. Specifically, we find a dividing curve in the space spanned by the dimensionless Strouhal number (quantifying unsteady inertia of the beam) and the compliance number (quantifying the strength of the fluid--structure coupling) separating scenarios of potential failure during transient conditions from potential failure at the steady load.
\end{abstract}


\end{frontmatter}



\section{Introduction}

In classical continuum mechanics, each material point is assumed to interact solely with its immediate neighbors \cite{truesdell1991first,gurtin1981introduction}. This \emph{local} interaction paradigm is typically captured through partial differential equations (PDEs) that utilize spatial derivatives. While this framework has proven successful in modeling a wide range of mechanical phenomena, it encounters significant challenges in at least two important scenarios. First, classical continuum theory struggles with discontinuities, where the field variables may be undefined or exhibit sharp gradients---such as during fracture, crack formation, and sudden material failure or separation \cite{silling2000}. Because the classical theory relies on being able to compute spatial derivatives of continuous fields, it lacks a straightforward means to handle abrupt damage within the material. Second, and related to the previous point, certain multiphysics phenomena are intrinsically \emph{nonlocal} \cite{voyiadjis2019handbook}, requiring knowledge of stresses, strains, or more generally the state of the body in the neighborhood around the material point under question (not just the nearest neighbor material points). Nonlocality can also be thought of as a manifestation of material responses controlled by internal length scales \cite{askari2008peridynamics}.

The theory of \emph{peridynamics} provides a robust alternative to classical continuum mechanics and avoids many of the limitations imposed by local spatial derivatives \cite{silling2000,silling2010}. Peridynamics is based on an integral equation of motion, making it inherently applicable regardless of whether discontinuities are present. Specifically, in three dimensions, the basic balance law of a peridynamic solid takes the general form
\begin{multline}
    \rho_s(\mathbf{x}) \frac{\partial^2 \mathbf{u}(\mathbf{x}, t)}{\partial t^2}
    \\
    =
    \iiint_{\mathcal{H}(\mathbf{x})} \mathbf{f} \Bigl(
    \mathbf{u}(\mathbf{x}', t) - \mathbf{u}(\mathbf{x}, t),
    \,\mathbf{x}' - \mathbf{x} \Bigr)
    \, d\mathcal{V}_{\mathbf{x}'} + \mathbf{b}(\mathbf{x}, t),
\label{eq:pd_balance}
\end{multline}
where $\rho_s$ is the mass density of the material, $\mathbf{u}$ is the displacement vector, and $\mathbf{b}$ is a body force.

In the formulation~\eqref{eq:pd_balance}, each point in the material, often referred to as the ``center particle'' $\mathbf{x}$, interacts with every other point $\mathbf{x}'$ within a finite region (neighborhood), $\mathcal{H}(\mathbf{x})$, surrounding $\mathbf{x}$. This finite region is known as the \emph{horizon} and is characterized by the horizon size $\delta$, an intrinsic length scale. The force functional $\mathbf{f}$ then maps the positions of the center particle and all particles within its horizon to the net force acting at $\mathbf{x}$. By relying on these integral interactions instead of spatial derivatives, peridynamics naturally accommodates cracks, damage, and other material discontinuities without compromising its mathematical validity \cite{silling2017why,silling2017introduction}. As a result, the theory of peridynamics has gained significant attention for modeling fracture initiation, propagation, and dynamic behavior in elastic materials \cite{silling2007peridynamic, silling2010,bobaru2015cracks,ha2010studies,silling2010crack}.

The interaction between a fluid flow and a compliant structure governs the performance and reliability of systems that span geophysical, biomedical, and micro-engineering scales. In the Earth's mantle, for example, channelized magma flow can weaken overlying lithospheric plates and initiate large-scale failure processes such as earthquakes and volcanic eruptions \cite{lithgow2004origin,steinberger2001large}, leading to the creation of \emph{laccolith} formations \cite{Michaut2011,Bunger2011,Bunger2011b,Thorey2014}. At the opposite length scale, organ-on-a-chip devices rely on a microfluidic perfusion layer to supply nutrients to delicate cell cultures; excessive shear or pressure in the flow damages the soft cell layer on the channel walls and compromises tissue viability \cite{mori2017skin,guenat2018incorporating}. Related concerns arise in microfluidic devices fabricated from polydimethylsiloxane (PDMS), where high-throughput flows can cause the thin PDMS ceilings of channels to fracture \cite{zhang2020experimental}. 

These examples motivate the need for a unified theoretical framework capable of predicting how compliant channel walls bend, yield, and ultimately fail due to internal viscous fluid flow. On the solid mechanics side, peridynamics has proved effective for thin structures---beams, plates, and membranes---capturing both large elastic deflections and the nucleation and growth of cracks (see, e.g., \cite{silling2005,ogrady2014a,ogrady2014b,yang2020,yang2022a,yang2022b,yang2023a,yang2023b,diyaroglu2015,naumenko2022,taylor2015} and the references therein). Its nonlocal integral form naturally accommodates the geometric discontinuities and damage evolution that can occur in a microchannel's ceiling. On the fluid mechanics side, the large aspect ratio of typical microchannels justifies the \emph{lubrication approximation}, which can be used to reduce the Navier--Stokes equations of hydrodynamics to a depth-averaged one-dimensional description involving only the pressure and flow rate, while retaining the essential features of viscous flow (see, e.g., \cite{panton2013incompressible,Stone2017}). Lubrication-based reduced-order models have therefore become a useful tool for coupling fluid motion with flexible beams, plates, shells, and membranes (see, e.g., \cite{Hosoi2004,Elbaz2014,Tulchinsky2016,Carlson2016,Boyko2019,inamdar2020unsteady,wang2022reduced} and the references therein).

This study couples a peridynamic formulation for a thin, compliant channel wall with a lubrication description of the underlying viscous flow. The resulting model captures both the nonlocal solid response and the fluid--structure interaction that drives deformation and damage. To this end, this study is organized as follows. Sec.~\ref{sec:prblmform} formulates the physical problem. Sec.~\ref{sec:goveqns} develops the dimensionless governing equations, introduces the key dimensionless groups---the horizon size $\Delta$, the Strouhal number $St$, the compliance number $\beta$, and the Reynolds number $Re$---and introduces the simulation methodology. Sec.~\ref{sec:results} discusses steady-state (Sec.~\ref{subsec:steady}) and transient (Sec.~\ref{subsec:DynamicSimulation}) simulations of the coupled fluid--peridynamic structure interaction problem. Sec.~\ref{subsec:dispersion} presents a dispersion study that traces how the key dimensionless numbers modify the phase velocity and spatial damping rate in this system. Then, Sec.~\ref{subsec:damage} compares the fracture `risk' between dynamic loading and an equivalent static-pressure loading, showing when and how the channel's ceiling may fail. Sec.~\ref{sec:limitations} discusses some of the advantages, as well as limitations, of our modeling approach, while Sec.~\ref{sec:conclusion} concludes our study and discusses avenues for future work.

\section{Problem formulation}
\label{sec:prblmform}

Following Inamdar et al.~\cite{inamdar2020unsteady}, we seek to address the fluid--structure interaction problem in which a two-dimensional (2D) channel with a top elastic wall of length $\ell$ is clamped at both ends. The top wall is modeled as a beam, with transverse (vertical) displacement $u(x,t)$. The channel's undeformed height is $h_0$. Thus, its deformed height is $h=h_0+u$. 

\begin{figure}[h]
    \centering
    \includegraphics[width=\columnwidth]{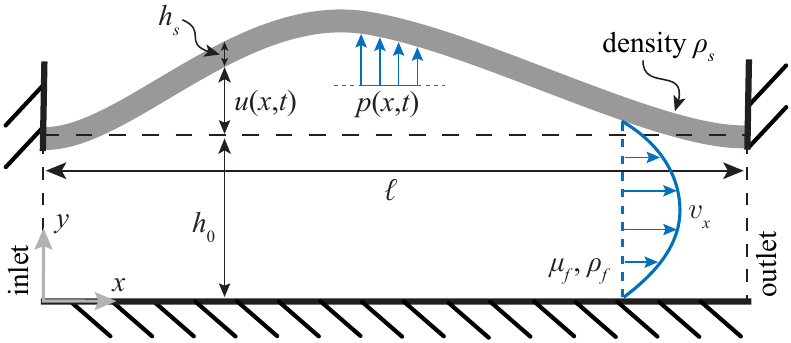}
    \caption{Schematic of the problem geometry and the mathematical notation.}
    \label{fig:prblmdef}
\end{figure}

The beam has a flexural rigidity denoted by $B$, a mass per area denoted by $m_s = \rho_s h_s$, and thickness $h_s$, as shown in Fig.~\ref{fig:prblmdef}. The flexural rigidity represents the beam's resistance to bending deformation and is given, per unit width, by $B=E_Y h_s^3/12$ for isotropic materials \cite{howell2009applied}, where $E_Y$ is the Young's modulus of the material. 

The Newtonian fluid in the channel below the beam has a dynamic viscosity $\mu_f$, density $\rho_f$, axial velocity component $v_x(x,y,t)$, and pressure field $p(x,t)$, as shown in the figure. The bottom wall ($y=0$) is rigid. The fluid exerts a pressure load $p$ on the beam, taking a positive value if the load is applied in the positive $y$-direction.

We take $x$ to be the coordinate of the material points along the neutral axis of the beam, and let $y$ denote the direction of the beam's deflection from the neutral axis when subjected to a load. For convenience, however, the origin is placed at the bottom-left corner (inlet) of the fluid channel on the rigid wall. Time is denoted by $t$. 

\section{Governing equations}
\label{sec:goveqns}
\subsection{Peridynamic beam theory}

The transverse displacements of a thin beam under bending can be calculated using Euler--Bernoulli beam theory \cite{howell2009applied}. This simplified version of the linear theory of elasticity provides a framework for analyzing the deflection behavior of beams, when the external loads applied perpendicular to the beam's longitudinal axis cause only small displacements and strains \cite{reddy2007theory}. 

To motivate the governing equation for the beam, let us first review some basics. Based on the Euler--Bernoulli beam kinematics, the strain tensor $\varepsilon_{ij}$ within the beam can be represented as
\begin{equation}\label{eq:strain}
    \varepsilon_{ij} =
    \begin{bmatrix}
    -y \displaystyle \frac{\partial^2 u}{\partial x^2} & 0 \\[4mm]
    0 & 0
    \end{bmatrix}.
\end{equation}
For a homogeneous isotropic linear elastic material, the stress tensor $\sigma_{ij}$ corresponding to the strain field in Eq.~\eqref{eq:strain} can then be represented as
\begin{equation}\label{eq:stress}
    \sigma_{ij} =
    \begin{bmatrix}
    -E_Y y \displaystyle \frac{\partial^2 u}{\partial x^2} & 0 \\[4mm]
    0 & 0
    \end{bmatrix}.
\end{equation}
Thus, employing summation convention, the strain energy density of the beam under classical continuum theory (CCM) is
\begin{equation}
    W_{\mathrm{CCM}} = \frac{1}{2} \varepsilon_{ij} \sigma_{ij} = \frac{E_Y}{2} y^2 \left( \frac{\partial^2 u}{\partial x^2} \right)^2.
    \label{eq:strain_energy_ccm}
\end{equation}

In the spirit of Eq.~\eqref{eq:pd_balance}, peridynamic beam theory concerns a nonlocal generalization of the Euler--Bernoulli beam based on a suitable nonlocal generalization of the curvature ${\partial^2 u}/{\partial x^2}$. Specifically, Yang et al.~\cite{yang2020,yang2023a,yang2022a} motivated the following version of Eq.~\eqref{eq:strain_energy_ccm} under peridynamic (PD) theory:
\begin{equation}
    W_{\mathrm{PD}} = \frac{E_Y}{2} y^2 \left[ \frac{1}{\delta} \int_{-\delta}^{+\delta} \frac{u(x +  \xi,t) - u(x,t)}{\xi^2} \, d\xi \right]^2,
    \label{eq:strain_energy_pd}
\end{equation}
where $\delta$ defines the size of the region of nonlocality, which is the horizon size in the context of peridynamics. It is expected that as $\delta \to 0^+$, the peridynamic beam theory should reduce to the classical beam theory. Indeed, performing Taylor series expansion of $u$ in $x$, and neglecting higher-order terms, it is straightforward to show that Eq.~\eqref{eq:strain_energy_pd} reduces to Eq.~\eqref{eq:strain_energy_ccm} as $\delta \to 0^+$. Then, Yang et al.~\cite{yang2020,yang2023a,yang2022a} followed an energy‐based methodology, in which the mechanical system's behavior is described by a Lagrangian. Employing the Euler--Lagrange equations to extremize the action, they obtained the equation of motion of the transverse displacement of a loaded peridynamic beam based on the strain energy density from Eq.~\eqref{eq:strain_energy_pd}. Specifically, 
\begin{multline}
    m_s \frac{\partial^2 u}{\partial t^2} - \frac{B}{\delta^2} \int_{-\delta}^{+\delta} \frac{1}{\xi^2}
    \left[ \int_{-\delta}^{+\delta} \frac{u(x + \eta,t) - u(x,t)}{\eta^2} \, d\eta \right. \\ \left.
    - \int_{-\delta}^{+\delta} \frac{u(x + \xi + \eta,t) - u(x + \xi,t)}{\eta^2} \, d\eta \right] d\xi = p(x,t).
\label{eq:pdbeam}
\end{multline}

In our problem, both ends of the beam are assumed to be clamped (see Fig.~\ref{fig:prblmdef}), thereby restricting two degrees of freedom at each boundary: the displacement $u$ and the rotation (which is the slope, $\partial u/\partial x$, in the CCM) \cite{reddy2007theory}. Thus, in the classical Euler--Bernoulli beam theory, clamping corresponds to the boundary conditions of $u=0$ and $\partial u/\partial x=0$ at $x=0$ and $x=\ell$. In the PD beam implementation, however, the rotation constraint is imposed using a nonlocal boundary treatment (specifically, by fictitious/ghost material points and symmetry conditions) rather than conditions on spatial derivatives. Therefore, following \citet{yang2020}, we formulate the following boundary conditions:
\begin{subequations}\begin{empheq}[left=\empheqlbrace]{alignat=3}
u(0,t) &= 0,    \quad   & u(-\zeta,t)      &= u(\zeta,t), \qquad &\zeta\in(0,2\delta], \\
u(\ell,t) &= 0,\quad   & u(\ell+\zeta,t)  &= u(\ell-\zeta,t), \qquad &\zeta\in(0,2\delta].
\end{empheq}\label{eq:ubc_dim}\end{subequations}
Here, the zero Dirichlet conditions fix the beam's displacement at the two ends. The mirror-image conditions enforce zero rotation in a PD-consistent manner by prescribing the displacement field in a fictitious boundary layer adjacent to each end. We note that there are other PD beam theories that enforce boundary conditions by treating the rotation as an independent kinematic variable \cite{naumenko2025first,yang2020peridynamic}.

\subsection{Lubrication theory incorporating flow inertia}

The 2D fluid flow, with velocity field $(v_x,v_y)$ and pressure field $p$, in the channel is governed by the continuity and Navier--Stokes equations for an incompressible Newtonian viscous fluid \cite{panton2013incompressible}:
\begin{subequations}\begin{empheq}[left=\empheqlbrace]{align}
    \frac{\partial v_x}{\partial x} + \frac{\partial v_y}{\partial y} &= 0,
    \label{eq:continuity}\\
    \frac{\partial v_x}{\partial t} + v_x \frac{\partial v_x}{\partial x} + v_y \frac{\partial v_x}{\partial y} 
    &= -\frac{1}{\rho_f} \frac{\partial p}{\partial x} + \frac{\mu_f}{\rho_f} \left( \frac{\partial^2 v_x}{\partial x^2} + \frac{\partial^2 v_x}{\partial y^2} \right),
    \label{eq:x-momentum}\\
    \frac{\partial v_y}{\partial t} + v_x \frac{\partial v_y}{\partial x} + v_y \frac{\partial v_y}{\partial y} 
    &= -\frac{1}{\rho_f} \frac{\partial p}{\partial y} + \frac{\mu_f}{\rho_f} \left( \frac{\partial^2 v_y}{\partial x^2} + \frac{\partial^2 v_y}{\partial y^2} \right).
    \label{eq:y-momentum}
\end{empheq}\label{eq:iNS}\end{subequations}
We are interested in a pressure-driven flow with a fixed inlet flow rate. Thus, at $x=0$, the area flow rate $\hat{q}$ (volumetric flow rate per unit width in this 2D problem) is imposed, while at $x=\ell$, the channel is considered open to the ambient and the pressure $p$ is set to zero gauge. Thus, the BCs on the flow are
\begin{equation}
    \hat{q}|_{x=0} = \hat{q}_0, \qquad p|_{x=\ell} = 0,
    \label{eq:flowbc_dim}
\end{equation}
for some constant $\hat{q}_0$.

We proceed by adopting the nondimensionalization of Inamdar et al.~\cite{inamdar2020unsteady} based on lubrication theory. To this end, we introduce the following dimensionless variables denoted by capital letters:
\begin{multline}
    H(X,T) = \frac{1}{h_0} h(x,t), \quad X = \frac{1}{\ell} x, \quad T = \frac{\hat{q}_0}{h_0 \ell} t, \quad Y = \frac{1}{h_0} y, \\
    \Delta = \frac{1}{\ell} \delta, \quad \Xi = \frac{1}{\ell} \xi, \quad \Theta = \frac{1}{\ell} \eta, \quad P(X,T) = \frac{h_0^3}{\mu_f \hat{q}_0 \ell} p(x,t), \\
    V_X(X,Y,T) = \frac{h_0}{\hat{q}_0} v_x(x,y,t), \quad V_Y(X,Y,T) = \frac{\ell}{\hat{q}_0} v_y(x,y,t).
    \label{eq:ndvars}
\end{multline}
Specifically, we nondimensionalize the channel height $h$ and the $y$-coordinate by the original (undeformed) channel height $h_0$. Meanwhile, the $x$-coordinate and all other longitudinal lengths are scaled by the axial length $\ell$ of the channel. The time scale and a characteristic streamwise velocity are derived from the imposed area flow rate $\hat{q}_0$. Then, by balancing the order of magnitude of terms in the dimensionless version of the continuity equation~\eqref{eq:continuity}, we obtain the characteristic velocity in the transverse direction \cite{panton2013incompressible}. Finally, the lubrication pressure scale is established by assuming that the order of magnitude of the axial pressure gradient, $\partial p / \partial x$, balances with that of the viscous term, $\mu_f \partial^2 v_x / \partial y^2$, in the $x$-momentum equation~\eqref{eq:x-momentum} \cite{panton2013incompressible}.

Next, we make the lubrication approximation, namely that the channel is long and thin: $h_0\ll\ell$. Then, we introduce the variables from Eq.~\eqref{eq:ndvars} into Eqs.~\eqref{eq:iNS} and neglect terms proportional to powers of $h_0/\ell\ll1$. Thus, the continuity equation and Navier--Stokes equations for the incompressible Newtonian fluid simplify to:
\begin{subequations}\begin{empheq}[left=\empheqlbrace]{align}
    \frac{\partial V_X}{\partial X} + \frac{\partial V_Y}{\partial Y} &= 0, \label{eq:continuity_nd_0}\\
    Re \left( \frac{\partial V_X}{\partial T} + V_X \frac{\partial V_X}{\partial X} + V_Y \frac{\partial V_X}{\partial Y} \right) &= -\frac{\partial P}{\partial X} + \frac{\partial^2 V_X}{\partial Y^2}, \label{eq:x-momentum_nd}\\
    \frac{\partial P}{\partial Y} &= 0, \label{eq:y-momentum_nd}
\end{empheq}\end{subequations}
Unlike the textbook approach \cite{panton2013incompressible}, we have retained terms multiplied by the \emph{effective} Reynolds number relevant to lubrication flows \cite{Stone2017},
\begin{equation}
    Re = \frac{h_0 \rho_f \hat{q}_0}{\ell \mu_f} = \frac{\text{flow inertia}}{\text{viscous forces}},
    \label{eq:Re}
\end{equation}
to account for weak flow inertia.

A no-slip BC is imposed on both the top and bottom walls of the channel. Since the top wall is in motion, the no-slip condition at the top wall also implies a kinematic (boundary) condition at $Y=H(X,T)$ that matches the vertical velocities of the flow and structure \cite{panton2013incompressible}. Enforcing the small-slope approximation inherent to the beam theory ($\partial H/\partial X\ll1$), the kinematic condition simplifies \cite{inamdar2020unsteady} to:
\begin{equation}
    \frac{\partial H}{\partial T} =  V_Y \big|_{Y=H} .
\label{eq:interfacebc}
\end{equation}
To derive a one-dimensional (1D) reduced model of the fluid--structure interaction, we integrate the continuity equation \eqref{eq:continuity_nd_0} from $Y=0$ to $Y=H(X,T)$. Defining the dimensionless flow rate as $Q(X,T) = \int_{0}^{H(X,T)} V_{X}(X,Y,T) \, dY$ and utilizing Eq.~\eqref{eq:interfacebc} to determine $V_{Y}|_{Y=H}$, we can immediately obtain the following 1D form of the continuity equation:
\begin{equation}
    \frac{\partial Q}{\partial X} + \frac{\partial H}{\partial T} = 0.
\label{eq:continuity_nd}
\end{equation}

Next, we also integrate the $X$-momentum equation~\eqref{eq:x-momentum_nd} from $Y=0$ to $Y=H(X,T)$. To do so, first, it is transformed into conservative form, then the horizontal velocity is assumed to follow a Hagen--Poiseuille (parabolic) flow profile: $V_{X} = {6Y(H-Y)Q}/{H^3}$, which is a von K\'arm\'an--Pohlhausen-type of closure  \cite{inamdar2020unsteady,stewart2009local,kalliadasis2012falling}. This closure is asymptotically consistent and valid at low Womersley number (i.e., for low flow oscillation frequency) \cite{Vosse2011,Zhang2024}, which is consistent with our time-independent flow BCs given in Eq.~\eqref{eq:flowbc_dim}. The pressure gradient is independent of $Y$ thanks to the $Y$-momentum equation~\eqref{eq:y-momentum_nd}. Performing these (lengthy but standard) calculations, we arrive at the 1D form of the axial momentum equation:
\begin{equation}
    Re\left[ \frac{\partial Q}{\partial T} + \frac{6}{5} \frac{\partial}{\partial X}\left(\frac{Q^2}{H}\right) \right] = -H \frac{\partial P}{\partial X} - \frac{12Q}{H^2}.
\label{eq:momentum_nd}
\end{equation}

\subsection{Summary of the model's governing equations}

Using the variables from Eq.~\eqref{eq:ndvars}, the peridynamic beam Eq.~\eqref{eq:pdbeam} can be made dimensionless as follows:
\begin{multline}
    St \frac{\partial^2 H}{\partial T^2} - \frac{1}{\Delta^2} \int_{-\Delta}^{+\Delta} \frac{1}{\Xi^2}
    \left[ \int_{-\Delta}^{+\Delta} \frac{H(X + \Theta,T) - H(X,T)}{\Theta^2} \, d\Theta \right. \\ 
    \left. - \int_{-\Delta}^{+\Delta} \frac{H(X + \Xi + \Theta,T) - H(X + \Xi,T)}{\Theta^2} \, d\Theta \right] d\Xi \\
    = \beta P(X,T).
\label{eq:pdbeam_nd}
\end{multline}
Here, two further dimensionless numbers emerge: the Strouhal number $St$ and the compliance number $\beta$ \cite{inamdar2020unsteady}, defined as:
\begin{align}
    St &= \frac{m_s \hat{q}_0^2 \ell^2}{B h_0^2} = \frac{(\text{solid's timescale for loading})^2} {(\text{fluid flow's timescale})^2} , \label{eq:St}\\
    \beta &= \frac{\mu_f \hat{q}_0 \ell^5}{B h_0^4} = \frac{\text{viscous flow forces}}{\text{beam bending forces}} .
    \label{eq:beta}
\end{align}
The Strouhal number $St$ quantifies the beam's unsteady inertia; a smaller $St$ corresponds to a faster loading response. Meanwhile, $\beta$ quantifies the fluid--structure interaction, combining the principal parameters that govern the fluid flow (viscosity and flow rate) and the structural deformation (bending rigidity); a smaller $\beta$ corresponds to weaker coupling between the flow and the structure. 
Finally, the dimensionless BCs from Eq.~\eqref{eq:ubc_dim} for the peridynamic beam governed by Eq.~\eqref{eq:pdbeam_nd}, corresponding to clamping at the ends, become:
\begin{subequations}\begin{empheq}[left=\empheqlbrace]{alignat=3}
    H(0, T) &= 1, \quad & H(-Z, T) &= H(Z, T), \qquad &Z \in (0, 2\Delta],\\
    H(1, T) &= 1, \quad & H(1+Z, T) &= H(1-Z, T), \qquad &Z \in (0, 2\Delta].
\end{empheq}\label{eq:Hbc_nd}\end{subequations}

The 1D lubrication model for the flow is governed by the integrated continuity~\eqref{eq:continuity_nd} and axial momentum equations~\eqref{eq:momentum_nd}, 
which are subject to the dimensionless inlet and outlet BCs based on Eq.~\eqref{eq:flowbc_dim}:
\begin{equation}
    Q(0,T) = 1, \qquad P(1,T) = 0.
    \label{eq:flowbc_nd}
\end{equation}

The initial conditions for Eqs.~\eqref{eq:pdbeam_nd}, \eqref{eq:continuity_nd} and \eqref{eq:momentum_nd} correspond to a state of rest:
\begin{equation}
    H(X,0) = 1,\qquad \frac{\partial H}{\partial T}(X,0) = 0, \qquad Q(X,0) = 0.
\end{equation}
No initial conditions are specified on the pressure as it is not an evolutionary variable in an incompressible flow. Note, however, that these initial conditions are rapidly ``forgotten'' as the coupled system evolves dynamically.

\subsection{Solution approach and parameter space}

In this study, we present the numerical solution of the one-dimensional fluid--peridynamic structure model composed of Eqs.~\eqref{eq:pdbeam_nd}, \eqref{eq:continuity_nd}, and \eqref{eq:momentum_nd} subject to the BCs in Eqs.~\eqref{eq:Hbc_nd} and \eqref{eq:flowbc_nd}. There are four dimensionless groups describing the dynamics, $Re$, $St$, $\beta$, and $\Delta$, as given in Eqs.~\eqref{eq:Re}, \eqref{eq:St}, and \eqref{eq:beta}.

The governing equations are discretized in space and time to determine the system's dynamic responses. The 1D axial domain is uniformly discretized, and the time integration is performed using the Newmark-$\beta$ time-stepping method \cite{subbaraj1989survey}. The technical details of the numerical methodology for the proposed 1D fluid--peridynamic structure model are given in \ref{subsec:NumericalScheme}. Briefly, we use a segregated iterative approach following \cite{inamdar2020unsteady,wang2022reduced}. We first solve the beam's motion equation~\eqref{eq:pdbeam_nd} given the pressure $P$ to obtain the deformed channel height $H$ and its time derivative ${\partial H}/{\partial T}$. Then, the flow rate $Q$ is found from the continuity Eq.~\eqref{eq:continuity_nd} by integrating ${\partial H}/{\partial T}$ over $X$. Finally, the pressure is computed from the axial momentum Eq.~\eqref{eq:momentum_nd} using second-order backward differences in time combined with central differences in space. These calculations are iterated until the solution converges at each time step.


\section{Results and discussion}
\label{sec:results}
\subsection{Steady state and validation}
\label{subsec:steady}

To validate our coupled fluid--peridynamic structure model, we compare the steady‐state solutions obtained in this work with those reported in the literature for a classical beam \cite{inamdar2020unsteady}. Given that the peridynamic beam theory approaches classical beam theory as the horizon size reduces (i.e., $\Delta \to 0^+$), we vary $\Delta$. Specifically, for the peridynamic formulation, we consider results for two different horizon sizes, $\Delta = 1/100$ and $\Delta = 1/20$. 

As the beam first responds to the flow, it undergoes a brief oscillatory transient before settling to a steady configuration. To shorten the transient time window in this steady-state benchmark, we reduced the Strouhal number $St$, thereby diminishing inertial effects and accelerating convergence. Steady state is considered achieved when the relative change in the displacement field between successive time steps, measured using the $L^2$ norm, falls below $10^{-5}$. Once steady state is achieved, we examine the beam's deformed shape and the corresponding pressure distribution along its length. We focus on the case of $\beta = 556$, and $Re = 0.5$ from \cite{inamdar2020unsteady}. 

\begin{figure}[ht!]
    \centering
    \begin{subfigure}[b]{\linewidth}
            \centering
            \includegraphics[width=0.7\textwidth]{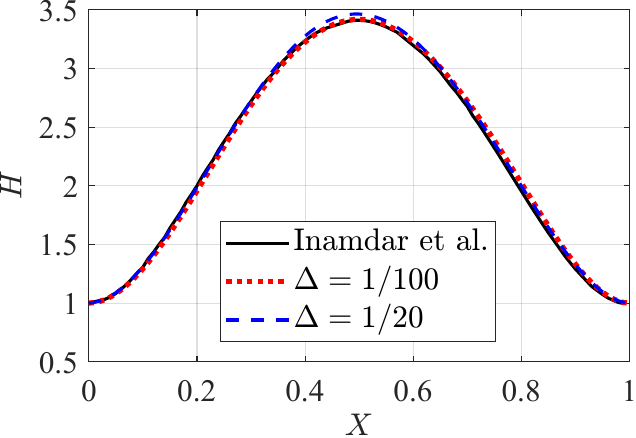} 
        \caption{ }
        \label{fig:compareH}
    \end{subfigure}
    \hfill 
    \begin{subfigure}[b]{\linewidth}
            \centering
            \includegraphics[width=0.7\textwidth]{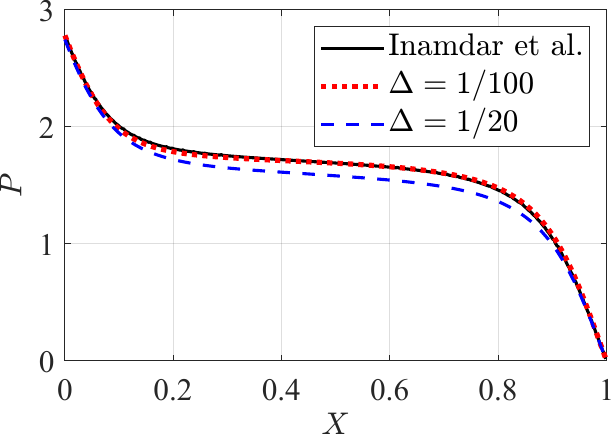}
        \caption{ }
        \label{fig:compareP}
    \end{subfigure}
    \caption{Steady state (a) deformed channel wall $H(X)$ and (b) pressure load $P(X)$ along the beam computed using classical continuum mechanics by Inamdar et al.~\cite{inamdar2020unsteady} (solid curves) and the fluid--peridynamic structure solver developed in this work (dashed curves), for three horizon sizes: $\Delta = 1/100$ and $\Delta = 1/20$. The simulation parameters are $\beta = 556$ and $Re = 0.5$.}
    \label{fig:comparison}
\end{figure}

Fig.~\ref{fig:comparison} compares our PD-based results with those obtained using CCM in \cite{inamdar2020unsteady}. As seen from the figure, both the beam deformation and the pressure distribution along the beam are in close agreement with the solution from the literature. Notably, reducing the horizon size yields an almost identical solution to the classical counterpart, indicating that the peridynamic theory is converging towards the classical one as the horizon size is decreased, as expected.

\subsection{Nonlinear dynamics and effect of the horizon}
\label{subsec:DynamicSimulation}

\begin{figure}[h]
    \centering
    \includegraphics[width=0.9\linewidth]{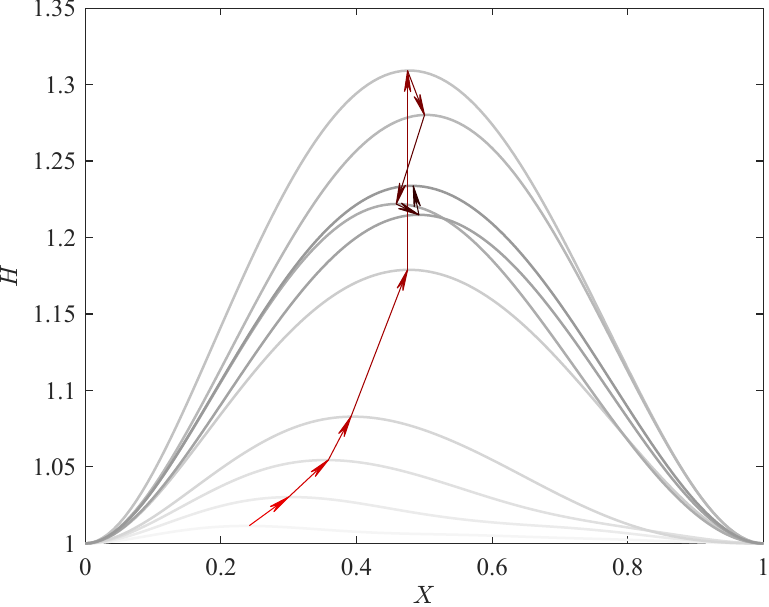}
    \caption{Nonlinear dynamics of the coupled problem illustrated by the time evolution of the channel height, $H(X,T)$. Darker curves represent later times, while lighter curves correspond to earlier times. The arrows indicate the direction of temporal evolution of the channel height. The simulation parameters are $St = 10$, $\beta = 20$, $Re = 0.5$, and $\Delta = 1/120$.}
    \label{fig:TvsH}
\end{figure}

Next, we discuss the dynamic simulations. A sample result is shown in Fig.~\ref{fig:TvsH}. We observe that, in general, the flow-induced deformation manifests as a bulge that originates near the flow inlet and migrates toward the middle of the channel. The main bulge-like deformation is accompanied by small wave-like perturbations. As the bulge travels downstream, the maximum height of the channel gradually increases. However, once the bulge reaches the middle region, the peak's height begins to fluctuate up and down. The bulge begins to oscillate laterally, shifting slightly from left to right around the centerline of the channel. These oscillations eventually subside as the system approaches the steady state, as discussed in Sec.~\ref{subsec:steady}. The frequency and amplitude of the vertical and horizontal oscillations are strongly influenced by the dimensionless parameters $St$, $\beta$, and $Re$, reflecting the nonlinear interplay between beam inertia, fluid--structure coupling, and flow inertia in this problem.

\begin{figure}[t]
    \centering
    \includegraphics[width=.9\linewidth]{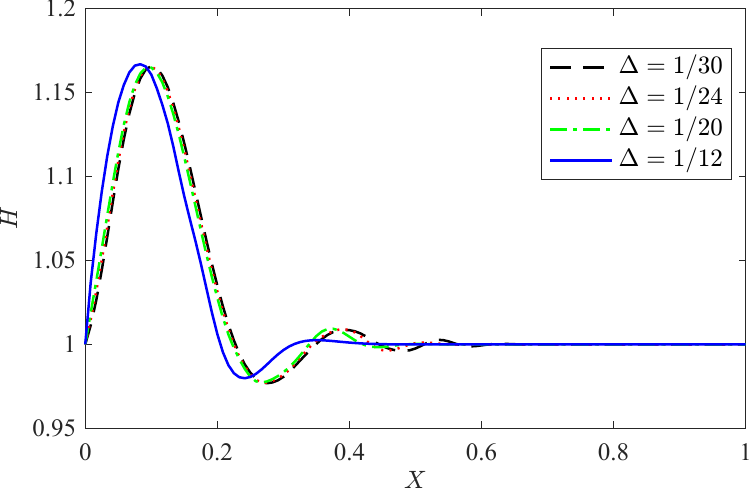}
    \caption{Effect of the horizon size $\Delta$ on the damping behavior of the channel height oscillations at an early time of $T = 0.02$, showing the stronger wave damping as $\Delta$ increases. The simulation parameters are $St = 10$, $\beta = 10^5$, and $Re = 10^{-5}$.}
    \label{fig:deltaeffect}
\end{figure}

The unsteady evolution is qualitatively similar to that in \cite{inamdar2020unsteady} under a CCM theory, as expected. To get a sense of the peridynamic aspect of the present problem, we next determine the influence of the dimensionless horizon length $\Delta$ on the dynamics. This influence is most easily observed at early times. Fig.~\ref{fig:deltaeffect} shows the deformed channel height recorded at an early time during the simulation ($T = 0.02$), for four different horizon sizes. The remaining parameters are fixed at  $St = 10$, $\beta = 10^5$, and $Re = 10^{-5}$ for these simulations.
From Fig.~\ref{fig:deltaeffect}, it is evident that the smaller oscillations to the right of the primary bulge undergo stronger damping when the horizon size is larger (e.g., $\Delta = 1/30$ and $1/24$ compared to $\Delta = 1/20$ and $1/12$). For $\Delta =  1/12$, these oscillations have nearly flattened, indicating more significant energy dissipation for stronger nonlocality and more effective dissipation of vibrational energy in the system.

Of course, the damping due to the nonlocality in this nonlinearly coupled problem depends strongly on the wavenumber of the oscillations, its amplitude, and the system parameters ($St$, $\beta$, and $Re$). Therefore, it is now of interest to determine the linearized dispersion relation \cite{tulchinsky2019frequency,eremeyev2025wave,eremeyev2025dispersion} of this coupled problem, to understand the damping quantitatively.

\subsection{Dispersion relation for the coupled problem}
\label{subsec:dispersion}

To linearize the system, we decompose the dimensionless variables into base states and small perturbations as
\begin{subequations}\begin{empheq}[left=\empheqlbrace]{alignat=3}
    H(X,T) &= H_0 + H'(X,T), \quad & H'(X,T) &= \hat{H}\,e^{i(kX-\omega T)},
\label{eq:H_exp}\\
    Q(X,T) &= Q_0 + Q'(X,T), \quad & Q'(X,T) &= \hat{Q}\,e^{i(kX-\omega T)},
\label{eq:Q_exp}\\
    P(X,T) &= P_0 + P'(X,T), \quad & P'(X,T) &= \hat{P}\,e^{i(kX-\omega T)},
\label{eq:P_exp}
\end{empheq}\label{eq:HQP_perturb}\end{subequations}
where $H_0$, $Q_0$, and $P_0$ denote the base state values, and $H'$, $Q'$, and $P'$ represent the corresponding perturbations.
Furthermore, to derive a linearized dispersion relation, we have assumed harmonic forms of the perturbations \cite{weckner2005,tulchinsky2019frequency,eremeyev2025wave,eremeyev2025dispersion} in Eqs.~\eqref{eq:HQP_perturb}. Here, $H'$, $P'$, and $Q'$ are understood as the \emph{real part} of the complex-valued expressions on the right-hand sides, but we leave the real part notation understood, as is common in the literature. In these expressions, $k$ is the (dimensionless) wavenumber, and $\omega$ is the (dimensionless) frequency. 

Now, we take the base state to be uniform, namely $H_0$, $Q_0$, and $P_0$ are constants, and we assume the perturbations to be small, namely $|\hat{H}| \ll |H_0|$, $|\hat{Q}| \ll |Q_0|$, $|\hat{P}| \ll |P_0|$, such that products of hat (and prime) quantities can be neglected. Under the harmonic assumption, the perturbation amplitudes $\hat{H}$, $\hat{Q}$, and $\hat{P}$ are also necessarily constants.

By substituting Eqs.~\eqref{eq:HQP_perturb} into the dimensionless peridynamic beam's equation of motion~\eqref{eq:pdbeam_nd} and the fluid's continuity and momentum Eqs.~\eqref{eq:continuity_nd} and \eqref{eq:momentum_nd}, respectively, and keeping only first-order terms in the perturbation amplitudes, we obtain:
\begin{subequations}\label{eq:linearized_eqs}%
\begin{align}
    \label{eq:pdbeam_nd_pert_harmonic}
    \left[ -St \, \omega^2  + \mathcal{D}(\Delta, k) \right] \hat{H} - \beta \hat{P} &= 0, \\
    \label{eq:continuity_nd_pert_harmonic}
    k \hat{Q} - \omega  \hat{H} &= 0,
\end{align}
and
\begin{multline}
    \left[Re \left( -i\omega + \frac{12}{5}\frac{ikQ_0}{H_0} \right) + \frac{12}{H_0^2}\right]\hat{Q} \\
    \qquad + \left(-ik Re \frac{6}{5}\frac{Q_0^2}{H_0^2} - 24\frac{Q_0}{H_0^3}\right)\hat{H} + ikH_0\hat{P} = 0,
    \label{eq:momentum_nd_pert_harmonic}
\end{multline}%
\end{subequations}%
where, for convenience, we have defined
\begin{equation}
    \mathcal{D}(\Delta, k) = \frac{1}{\Delta^2} \left( 2k\sum_{n=1}^{\infty}\frac{(-1)^n(k\Delta)^{2n-1}}{(2n-1)(2n)!} \right)^2.
    \label{eq:mathcal_D}
\end{equation}
Technical details of the linearization are given in \ref{subsec:Linearization}. 

An important consistency check of the nonlocal formulation is its reduction to classical beam theory as the horizon size vanishes (i.e., $\Delta \to 0^+$). Under classical beam theory, the perturbation obeys
\begin{equation}\label{eq:classical_beam}
    -St\, \omega^2 \hat{H} + k^4\hat{H} = \beta \hat{P}
\end{equation}
instead of Eq.~\eqref{eq:pdbeam_nd_pert_harmonic}. Indeed, by noting that the series in Eq.~\eqref{eq:mathcal_D} is alternating with terms whose absolute values decrease monotonically to zero, it can be shown that 
\begin{equation}
\begin{aligned}
\mathcal{D}(\Delta, k)
&= \frac{1}{\Delta^{2}}
   \left[ 2k\left(-\frac{k\Delta}{2}+O(\Delta^{3})\right)\right]^{2} \\[4pt]
&= \frac{1}{\Delta^{2}}
   \left[k^{4}\Delta^{2}+O(\Delta^{4})\right] \\[4pt]
&= k^{4}+O(\Delta^{2})
   \;\xrightarrow{\Delta\to 0}\;k^{4},
\end{aligned}
\end{equation}
thereby recovering the standard $k^4$ bending term in Eq.~\eqref{eq:classical_beam} as expected.

To identify the eigenmodes of the coupled fluid--peridynamic structure system, we focus on the following base state:
\begin{equation}\label{eq:base_state}
    H_0 = 1,\quad Q_0 = 1,\quad P_0 = 0.
\end{equation}
Here, to satisfy the BCs~\eqref{eq:Hbc_nd}--\eqref{eq:flowbc_nd}, the undeformed height and reference flow rate are normalized to unity, and the pressure is taken to be zero, without loss of generality.
Thus, the linearized problem consisting of Eqs.~\eqref{eq:linearized_eqs} for the base state in Eq.~\eqref{eq:base_state} can then be cast into a  matrix form:
\begin{equation}\label{eq:matrix_system}
    \begin{bmatrix}
    -St\,\omega^2 + \mathcal{D}(\Delta, k) & -\beta & 0 \\[2mm]
    -\omega & 0 & k \\[2mm]
    -\left( 24 + i\,\frac{6}{5} k Re \right) & ik & 12 + i \left( \frac{12}{5}k -\omega \right) Re
    \end{bmatrix}
    \begin{bmatrix}
    \hat{H} \\[2mm]
    \hat{P} \\[2mm]
    \hat{Q}
    \end{bmatrix}
    =
    \begin{bmatrix}
    0 \\[2mm]
    0 \\[2mm]
    0
    \end{bmatrix},
\end{equation}
where the first row arises from the PD beam equation, the second from the continuity equation, and the third row from the fluid's axial momentum balance equation.

A trivial solution (i.e., $\hat{H}=\hat{Q}=\hat{P}=0$) of Eq.~\eqref{eq:matrix_system} always exists. However, for a nontrivial solution to exist, the determinant of the coefficient matrix on the left-hand side of  Eq.~\eqref{eq:matrix_system} must vanish. This condition gives rise to the \emph{dispersion relation} $\omega(k)$, or $k(\omega)$. Specifically, $\omega$ or $k$ can be obtained as a function of the other by solving the algebraic equation
\begin{equation}\label{eq:dispersion_relation}
    a \omega^2 + b \omega + c = 0,
\end{equation}
where
\begin{subequations}\begin{align}
    \label{eq:a}
    a &= St\,k^2 + \beta\,Re,\\
    \label{eq:b}
    b &= \left(- \frac{12}{5} \beta \, Re \, k + i \, 12 \beta\right),\\
    \label{eq:c}
    c &= \frac{6}{5} k^{2} \beta \, Re - i \, 24 k\beta - \frac{k^2}{\Delta^2}\left( 2k\sum_{n=1}^{\infty} \frac{(-1)^n (k\Delta)^{2n-1}}{(2n-1)(2n)!} \right)^2.
\end{align}\end{subequations}

The dispersion relation in Eq.~\eqref{eq:dispersion_relation} can be solved to determine the dimensionless phase velocity, $v_{p} = \Re{(\omega)}/\Re{(k)}$. Unlike the case of an uncoupled peridynamic beam or bar~\cite{weckner2005,eremeyev2025wave,eremeyev2025dispersion}, where the dispersion relation depends solely on the wavenumber, the present model exhibits a more complex behavior: the dispersion relation also depends on the key dimensionless parameters, $St$, $\beta$, and $Re$.

\subsubsection{Structural modes, \texorpdfstring{$\beta = 0$}{beta = 0}}
\label{subsec:pure_structure_dispersion}

\begin{figure}[ht!]
  \centering
  \includegraphics[width=.8\linewidth]{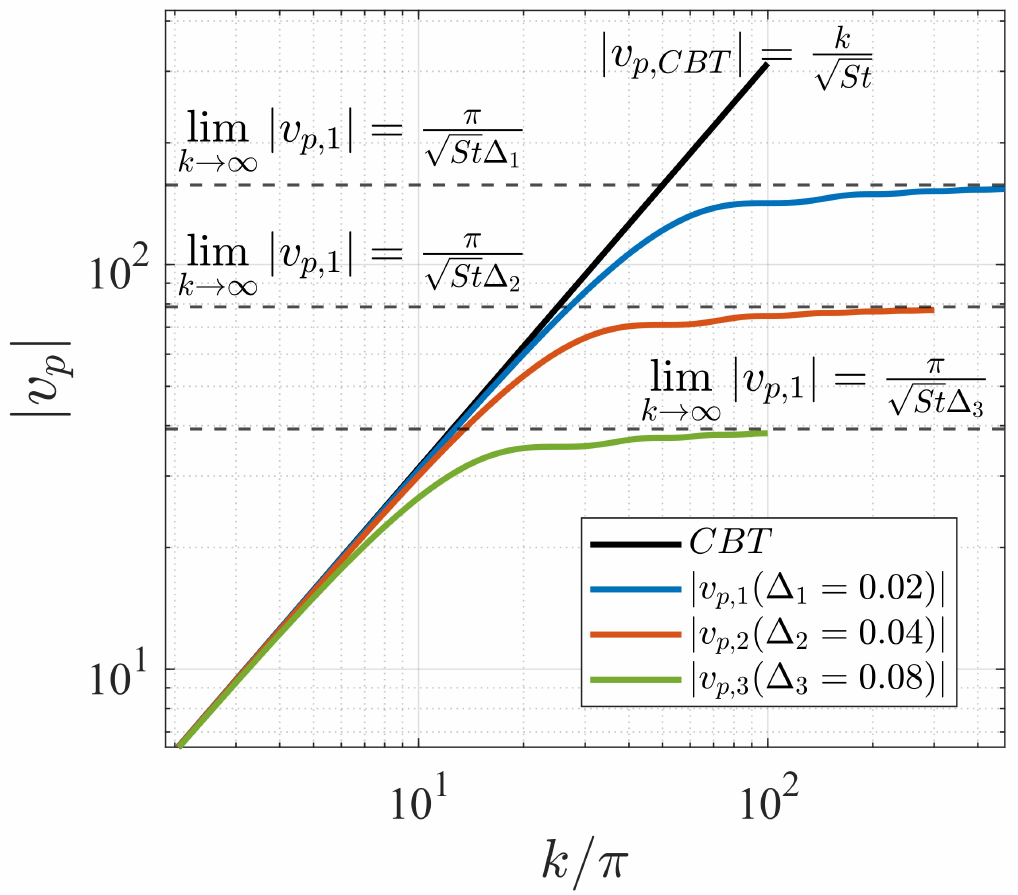}
  \caption{Dimensionless phase velocity $|v_{p}|$ versus $k/\pi$ for the classical Euler--Bernoulli beam (CBT) and the nonlocal PD beam with three horizon lengths $\Delta_{1}=0.02$, $\Delta_{2}=0.04$, and $\Delta_{3}=0.08$; $\beta=0$.  Dashed horizontal lines show the large-$k$ asymptotes given in Eq.~\eqref{eq:k_limit}.}
  \label{fig:dispersion1}
\end{figure}

Fig.~\ref{fig:dispersion1} shows the magnitude of the dimensionless phase velocity, $|v_{p}|$, for the classical Euler--Bernoulli beam (referred to as ``classical beam theory'' or ``CBT'') and for its nonlocal counterpart at \emph{zero fluid damping} ($\beta = 0$). The nonlocal attenuation visible at large $k$ is therefore due solely to the nonlocality (integral kernel); no viscous damping is present in these curves. The black curve ($\Delta = 0$) reproduces the CBT result, $|v_{p}| = k/\sqrt{St}$, and serves as the baseline. All nonlocal curves coincide with this linear behavior for long waves ($k\Delta \ll 1$), confirming that nonlocality leaves the low-frequency response unchanged. For a finite horizon, the curves diverge once $k\Delta = O(1)$, and each nonlocal branch levels off to the limiting value
\begin{equation}
  \lim_{k\to\infty} |v_{p}| = \frac{\pi}{\sqrt{St}\,\Delta}.
  \label{eq:k_limit}
\end{equation}

The dispersion curves in Fig.~\ref{fig:dispersion1} are governed by two independent parameters.  The Strouhal number $St$ comes from the beam inertia and fixes the reference slope: the CBT limit ($\Delta=0$) gives $|v_{p}| = k/\sqrt{St}$, so increasing $St$ shifts all velocities downward in proportion to $1/\sqrt{St}$. The second parameter, the peridynamic horizon $\Delta$, controls the degree of nonlocality. For any nonzero $\Delta$, the curve departs from the classical line once $k\Delta=O(1)$ and gradually saturates to the limiting value given in Eq.~\eqref{eq:k_limit}. Hence, a larger horizon lowers the high-$k$ plateau, demonstrating the intrinsic high-frequency filtering of the nonlocal kernel, whereas a larger $St$ lowers the entire curve without altering its shape.

\begin{figure}[ht!]
  \centering
  \includegraphics[width=.8\linewidth]{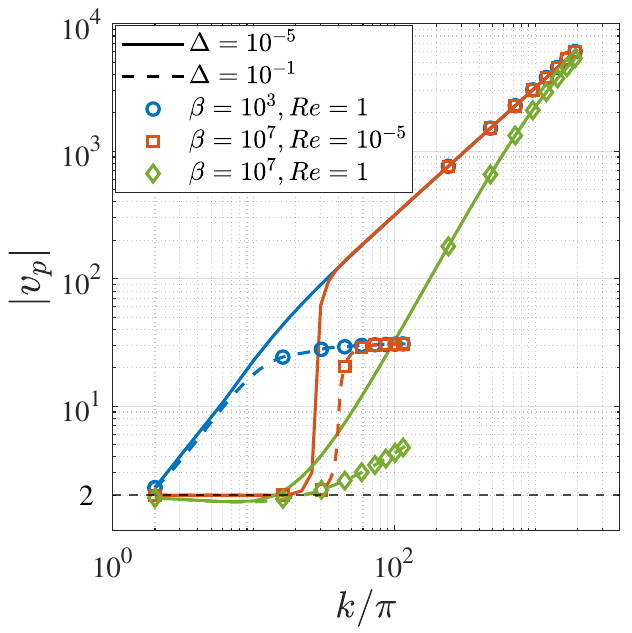}
  \caption{Dimensionless phase velocity $|v_{p}|$ versus $k/\pi$ for two horizons
           ($\Delta = 10^{-5}$, solid; $\Delta = 10^{-1}$, dashed)
           and three model parameter sets:
           $\beta=10^{3}$, $Re=1$ ({\color{NavyBlue}$\bm\circ$});
           $\beta=10^{7}$, $Re=10^{-5}$ ({\color{BrickRed}$\bm\square$});
           $\beta=10^{7}$, $Re=1$ ({\color{LimeGreen}$\bm\lozenge$}).
           The horizontal black dashed line marks the long-wave $\beta$-limit,
           $|v_{p}|=2$.}
  \label{fig:betaReDisp}
\end{figure}

This saturation persists for a significant range of the compliance numbers ($\beta<10^{3}$) and for finite Reynolds number $Re$, confirming that the elastic nonlocal kernel (not fluid damping) governs the short-wave response. When $\beta$ is small, the coupling is weak, and the dispersion becomes essentially insensitive to $Re$, provided the flow stays within the low-Reynolds-number regime relevant to microchannels, $Re\le1$. The break in slope at $k\Delta\approx1$ marks the onset of a ``wave-lag'' effect: beyond this point, short wavelengths travel progressively slower than the prediction from CBT. The same lag is evident in Fig.~\ref{fig:deltaeffect}, where waves recorded at a fixed instant arrive later as $\Delta$ increases. Altogether, the dispersion plot illustrates three features of the fluid--peridynamic structure interaction problem: agreement with classical beam theory at long wavelengths, attenuation of high-frequency waves, and a horizon- and $St$-dependent limit on the ultimate phase velocity.

\subsubsection{Fluid--structure coupled modes, \texorpdfstring{$\beta\ne0$}{beta != 0}}

Fig.~\ref{fig:betaReDisp} illustrates how fluid--structure coupling (quantified by $\beta$) and flow inertia (quantified by $Re$) modify the dispersion relation, keeping the intrinsic nonlocal filtering imposed by the horizon $\Delta$ fixed. As in the pure-structural case ($\beta=0$) from Sec.~\ref{subsec:pure_structure_dispersion}, the averaging over each horizon leads to high-spatial-frequency attenuation: for $k\Delta\gg1$, all curves converge to the universal plateau $\pi/(\sqrt{St}\,\Delta)$, irrespective of $\beta$ or $Re$. Increasing $\beta$ lowers the phase velocity of long-wave modes; in the limit of very strong coupling, the branch approaches the limiting value $|v_{p}|=2$. Thus, large $\beta$ suppresses the transmission of low-frequency waves without affecting their high-$k$ behavior. 

Between the low-$k$ plateau and the nonlocal asymptote, an \emph{intermediate} region appears whose shape depends on the Reynolds number. When $Re = O(1)$, the phase velocity grows quadratically throughout this interval before merging smoothly into the high-$k$ branch. In contrast, for $Re \ll 1$, the curve features an abrupt rise at the onset of the intermediate band and then relaxes monotonically toward the same asymptotic trend. Hence, we have demonstrated that in this coupled fluid--peridynamic structure model, $\beta$ controls the suppressed low-$k$ level, $\Delta$ fixes the high-$k$ limit, and $Re$ determines how the dispersion curve transitions between the two regimes.

\subsubsection{Damping}

\begin{figure}[ht]
  \centering
  \includegraphics[width=\linewidth]{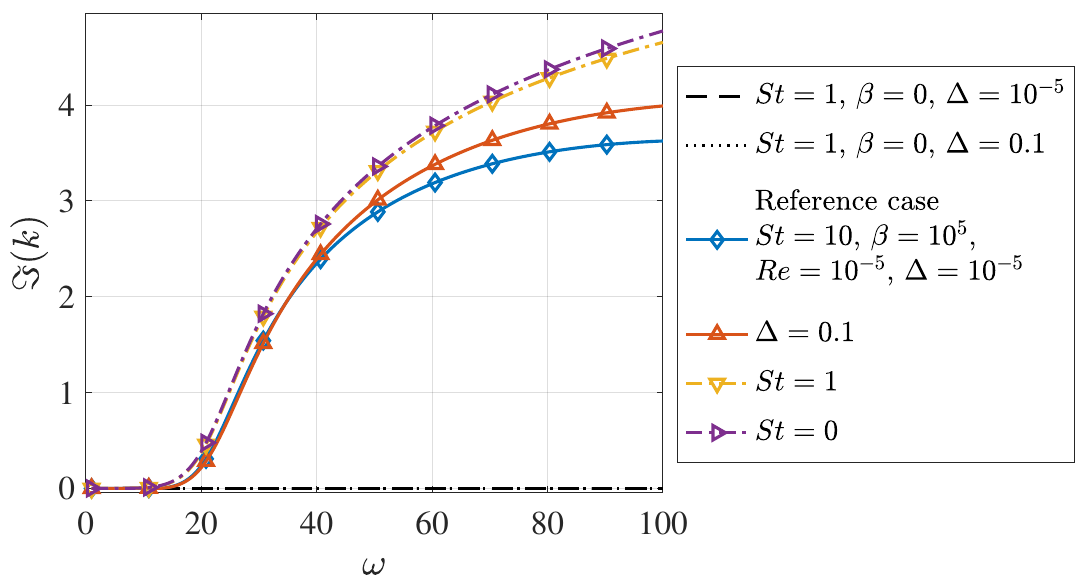}
  \caption{Spatial damping rate
         $\Im\bigl(k\bigr)$ as a function of $\omega$ for several sets of parameters.
         Dashed and dotted black curves: uncoupled beam
         ($\beta=0$) with very weakly nonlocal ($\Delta=10^{-5}$) and
         nonlocal ($\Delta=0.1$) elasticity, respectively.
         {\color{NavyBlue}$\bm\lozenge$}: reference coupled case
         $St=10$, $\beta=10^{5}$, $Re=10^{-5}$, $\Delta=10^{-5}$.
         {\color{RedOrange}$\bm\triangle$}: identical parameters except
         $\Delta=0.1$.
         {\color{Goldenrod}$\bm\triangledown$} and {\color{Purple}$\bm\triangleright$}: as in the reference case but with $St=1$ and $St=0$, respectively.}
  \label{fig:damping1}
\end{figure}

Next, Fig.~\ref{fig:damping1} distinguishes the roles of fluid--structure coupling, peridynamic horizon, and beam inertia in determining the spatial decay rate (damping) of harmonic waves. In the absence of coupling to the flow ($\beta = 0$), the attenuation is identically zero for both the local and nonlocal beams, confirming that a purely elastic solid supports loss-free wave propagation, as expected. Introducing fluid--structure coupling ($\beta=10^{5}$) generates a finite, frequency-dependent attenuation even for the practically local case (infinitesimal horizon size). 

As Fig.~\ref{fig:damping1} shows, the attenuation is negligible at small $\omega$, but rises sharply once the driving frequency reaches the point where fluid--structure coupling becomes effective. Subsequently, the attenuation reaches a broad plateau characteristic of fluid-dominated motion. Enlarging the horizon to $\Delta=0.1$ raises the entire reference curve, indicating that, while nonlocality supplies an additional (secondary) source of damping, viscous coupling remains the dominant mechanism in this regime. 

Similarly, reducing $St$ from ten to unity and then to zero can also slightly elevate the attenuation produced by fluid--structure coupling. A finite horizon, therefore, renders the structure effectively stiffer than its local (classical) counterpart. Indeed, the dynamic simulation shown in Fig.~\ref{fig:deltaeffect} above confirms this trend: increasing $\Delta$ produces visibly damped oscillations, demonstrating how nonlocality augments the overall spatial damping.

\begin{figure}[ht]
  \centering
  \includegraphics[width=0.9\linewidth]{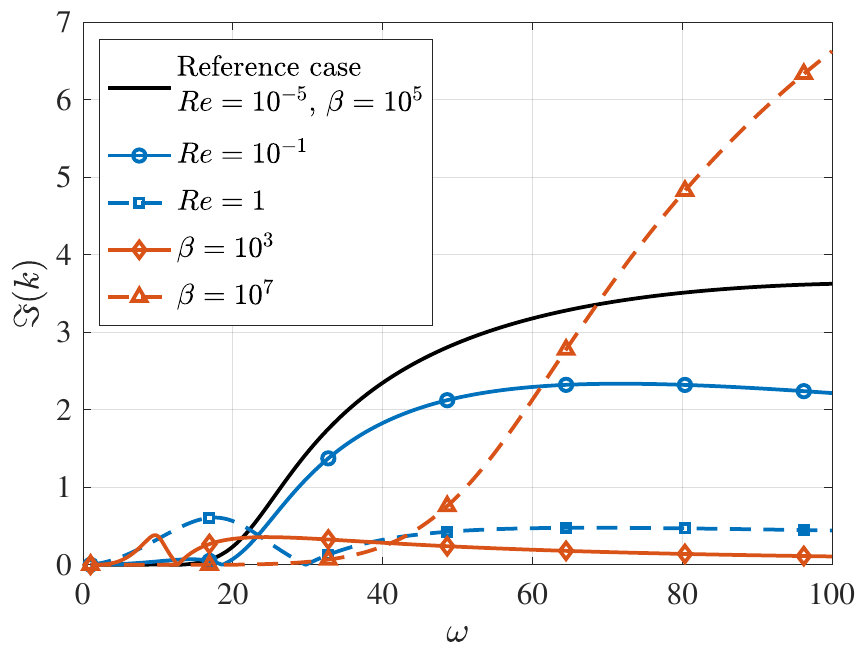}
  \caption{Spatial damping rate
           $\Im\bigl(k\bigr)$ as a function of $\omega$ at fixed Strouhal
           number $St = 10$ and vanishing horizon $\Delta = 10^{-5}$.  
           Reference case (solid black):
           $Re=10^{-5}$, $\beta = 10^{5}$.  
           Effect of $Re$ (blue):
           $Re=10^{-1}$ ({\color{NavyBlue}$\bm\circ$}) and
           $Re=1$ ({\color{NavyBlue}$\bm\square$}).  
           Effect of $\beta$ (orange):
           $\beta = 10^{3}$ ({\color{RedOrange}$\bm\lozenge$}) and
           $\beta = 10^{7}$ ({\color{RedOrange}$\bm\triangle$}).}
  \label{fig:dampingReBeta}
\end{figure}

Finally, Fig.~\ref{fig:dampingReBeta} isolates the influence of the Reynolds number $Re$ and the compliance number $\beta$ on the spatial damping rate of a beam with very weak nonlocality ($\Delta=10^{-5}$).  Relative to the reference case ($Re=10^{-5}$, $\beta=10^{5}$), decreasing $Re$ or increasing $\beta$ reduces the attenuation in the low-frequency range but enhances it at high frequencies. In other words, a highly viscous flow or a strong fluid--structure coupling extracts little energy from slowly varying waves, but dissipates more rapidly once the oscillation rate is high enough for viscous stresses to dominate.

Each curve intersects the $\Im(k)=0$ axis at isolated frequencies. These zeros correspond to eigenfrequencies at which the net power transferred between fluid and structure vanishes, allowing waves to propagate without spatial decay, despite the presence of damping mechanisms. As the coupling strength increases (larger $\beta$) or the viscous effects weaken (smaller $Re$), the first loss-free frequency shifts to higher values.

\subsection{Damage and failure of the microchannel: Static vs dynamic loading}
\label{subsec:damage}

Classical stress-based failure theories state that a linearly elastic beam fails when the axial stress on its outermost layer ($y=h_s$) exceeds the material strength, $\sigma_{\text{cr}}$ (equal to the ultimate strength $\sigma_{u}$ for brittle materials or to the yield strength $\sigma_{y}$ for ductile ones) \cite{goodno2018mechanics}. For an Euler--Bernoulli beam, based on Eq.~\eqref{eq:stress}, this criterion reads
\begin{equation}
\left|\frac{E_Y\,h_s}{2}\,
        \frac{\partial^{2}u}{\partial x^{2}}\right|>\sigma_{\text{cr}}.
\label{eq:stress_failure}
\end{equation}

Rather than prescribing a numerical value of $\sigma_{\mathrm{cr}}$, we recast Eq.~\eqref{eq:stress_failure} as an \emph{intrinsic, dimensionless curvature threshold}, $C_{\mathrm{cr}}$. To this end, let us define the (dimensionless) nonlocal curvature, $C(\mathbf{x},\mathbf{x}')$, of a bond \(\mathbf{x}'-\mathbf{x}\) as the average of the curvatures at the two particles that are connected by the bond $\mathbf{x}'-\mathbf{x}$, namely:
\begin{equation}
\begin{aligned}
C(\mathbf{x},\mathbf{x}')
     &=\frac{\left|C(\mathbf{x})\right|+\left|C(\mathbf{x}')\right|}{2}, \\
C(\mathbf{x})
     &=\frac{1}{\Delta}
        \int_{-\Delta}^{+\Delta}
        \frac{H(X(\mathbf{x})+\Theta,T)-H(X(\mathbf{x}),T)}
             {\Theta^{2}}\,
        d\Theta,
\end{aligned}
\label{eq:bond_curvature}
\end{equation}
Note that $C(\mathbf{x})$ depends implicitly on $X$ and $T$, i.e., $C(\mathbf{x}) = C(\mathbf{x};X,T)$, but we suppress this notation for clarity. $C(\mathbf{x})$ is computed nonlocally by integrating over the horizon surrounding the central particle $\mathbf{x}$, which ensures a consistent representation of the curvature within the peridynamic framework, recall Eq.~\eqref{eq:strain_energy_pd}. Thus, finally, the stress criterion becomes
\begin{equation}
     C(\mathbf{x},\mathbf{x}')
     > C_{\text{cr}}, 
\qquad
     C_{\text{cr}}
     = \frac{2\ell^{2}\,\sigma_{\text{cr}}}{E_Y\,h_0\,h_s}.
\label{eq:curv_threshold}
\end{equation}

In our simulations, we are only interested in the incipience of failure. Thus, $C(\mathbf{x})$ is evaluated everywhere, but \emph{no bonds are broken and no cracks are propagated}. $C_{\mathrm{cr}}$ is retained symbolically because no specific $\sigma_{\mathrm{cr}}$ is selected, so the numerical results depend only on the beam kinematics and material constants, which are already fixed in the model.
The maximum bond curvature,
\begin{equation}
    C_{\max} = \max_{\mathbf{x},\mathbf{x}'} C(\mathbf{x},\mathbf{x}'),
\end{equation}
is reported for each load case.  
Once a material strength is specified, the inequality in Eq.~\eqref{eq:curv_threshold} can be applied \emph{post hoc} to classify the outcome:
\begin{itemize}
  \item If $C_{\max} < C_{\mathrm{cr}}$: the beam would survive the load.
  \item If $C_{\max} > C_{\mathrm{cr}}$: the beam would fail.
  \item If $C_{\mathrm{cr}}$ lies between two reported values of $C_{\max}$: one load case would fail, the other would not.
\end{itemize}
This procedure cleanly separates the kinematic question (\emph{How large can the curvature become?}) from the material question (\emph{Where is the failure threshold?}).

\begin{figure}[ht]
    \centering
    \includegraphics[width=0.7\columnwidth]{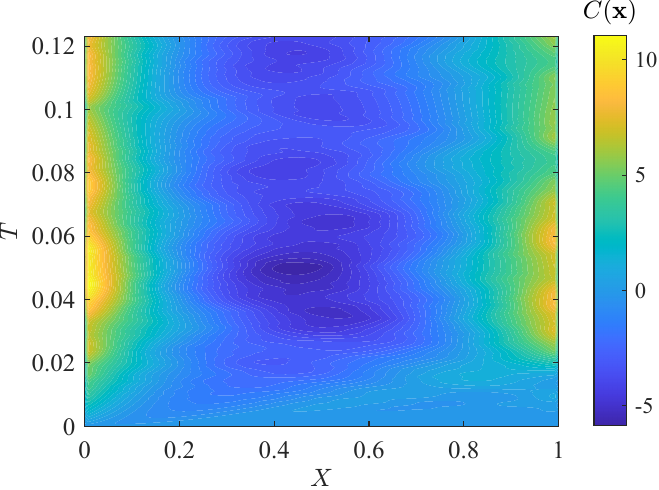}
    \caption{Contour plot showing the space-time evolution of the particle curvature field $C(\mathbf{x})$ as a function of dimensionless axial position $X\in[0,1]$ and dimensionless time $T\in[0,0.12]$. Colors show how $C(\mathbf{x})$ evolves over the chosen space-time interval, with warmer colors (yellow) indicating higher values and cooler colors (blue) indicating lower values. The superimposed contours indicate constant $C$-values. Simulation parameters are $St = 10$, $\beta = 20$, $Re = 0.5$, and $\Delta = 1/120$.}
    \label{fig:TvsooR}
\end{figure}

Fig.~\ref{fig:TvsooR} presents a contour space-time plot depicting the evolution of $C(\mathbf{x})$. The results indicate that the maximum curvature consistently appears at both ends of the clamped-clamped wall exposed to the viscous fluid flow.  The region near the inlet exhibits the larger curvature, which suggests that failure is most likely to occur near the inlet (if it does). Oscillations in the curvature are also observed, corresponding to the oscillations in the channel height $H$ seen in Fig.~\ref{fig:TvsH}.

\begin{figure}[ht!]
    \centering
    \includegraphics[width=\columnwidth]{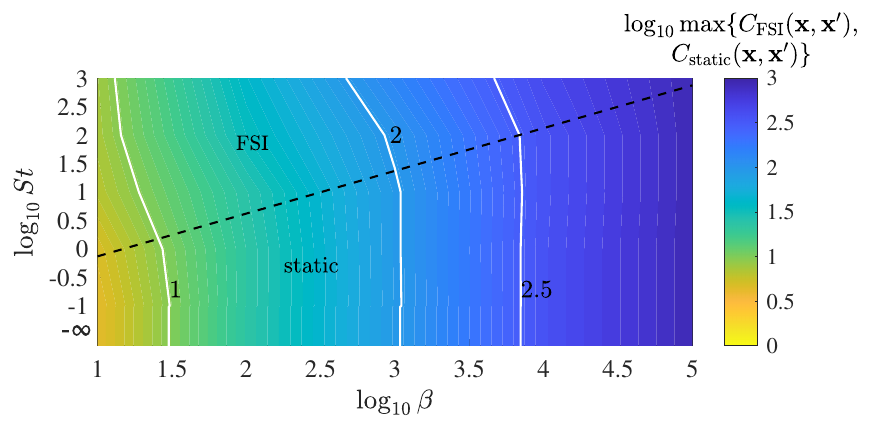}
    \caption{Contour plot illustrating the maximum dimensionless bond curvature across the dynamic (``FSI'') and static loading scenarios, with $\beta$ on the horizontal axis and $St$ on the vertical axis, while $Re$ is fixed to $0.5$. The dashed line, given by the fit in Eq.~\eqref{eq:dividing_line}, divides the parameter space into two regimes of (potential) beam failure. Note the logarithmic axes.}
    \label{fig:FSIvsStatic}
\end{figure}

Two loading scenarios are investigated:
\begin{enumerate}
  \item \textbf{Transient loading}: the beam is driven by the time-dependent pressure field produced by the coupled unsteady viscous flow.
  \item \textbf{Equivalent static loading}: the beam is subjected to static loading based on the steady-state pressure distribution extracted from the same fluid--structure interaction (FSI) simulation.
\end{enumerate}
We explored the parameter space spanned by the two key dimensionless groups: $St$ and $\beta$. To illustrate how these parameters affect the beam deformation and potential failure, we created contour plots that show the maximum dimensionless bond curvature across both scenarios. These plots map $\beta$ on the horizontal axis to $St$ on the vertical axis, while $Re$ is held fixed. The color scale indicates the level of bond curvature, highlighting how changes in $\beta$ and $St$ affect the structural response. We show an example of such a plot for $Re = 0.5$ in Fig.~\ref{fig:FSIvsStatic}. 

Furthermore, Fig.~\ref{fig:FSIvsStatic} shows that the $(St,\beta)$ plane is divided by a line in the log-log plot,
\begin{equation}\label{eq:dividing_line}
  \log_{10}{St} = 0.75\,\log_{10}\beta - 0.88,
\end{equation}
which separates the \emph{transient‐dominated} from the \emph{static‐dominated} failure modes. Parameter sets lying \textit{above} this line (i.e., cases with $St/\beta$ large) reach their maximum dimensionless bond curvature during the unsteady stage of the loading, so that $\max C_\mathrm{FSI} > \max C_\mathrm{static}$. The larger transient peak is generated by the pronounced up-and-down motion of the main deformation bulge as it moves away from
the inlet (recall the discussion of Fig.~\ref{fig:TvsH}). If a critical curvature is chosen such that
\begin{equation}
  \max C_\mathrm{static} < C_{\mathrm{cr}} < \max C_\mathrm{FSI},
\end{equation}
the beam would fail under the transient (dynamic) load, but would
remain safe under the equivalent static load.

Below the dividing line (i.e., for $St/\beta$ small), the ordering reverses, $\max C_\mathrm{static} > \max C_\mathrm{FSI}$, implying that the greatest curvature is attained only after the flow transients have settled. Hence, for all parameter pairs beneath the boundary, the curvature recorded in the steady state controls failure: the beam fails if $C_{\mathrm{cr}} < \max C_\mathrm{static}$ and survives otherwise.

This outcome highlights how transient (dynamic)forces in the coupled problem can significantly amplify beam deformation, while also revealing a clear boundary between dynamic and static (potential) failure mechanisms, enabled by the proposed reduced-order fluid--peridynamic structure interaction model.

We also note that varying the peridynamic horizon $\delta$ modifies the nonlocal structural response, and thus both the quasi-static and dynamic deflection/curvature fields can vary with the dimensionless horizon~$\Delta$. Consequently, even when the critical curvature threshold $C_{\mathrm{cr}}$ is held fixed, the predicted damage initiation site and onset time (i.e., the first location and time at which the computed curvature reaches $C_{\mathrm{cr}}$) may shift with $\Delta$. In the parameter regime explored here, these shifts are modest and are therefore not a primary focus of the present study.

\section{Advantages and limitations of the proposed modeling approach}
\label{sec:limitations}

The present work couples a one-dimensional peridynamic Euler--Bernoulli beam model (used over the entire structural span) with a classical lubrication-theory model for the fluid. The coupling is multiphysics: the fluid enters the beam equation through a distributed transverse pressure, while the lubrication equation is closed through the gap-height kinematics implied by the beam deflection. Importantly, we do not perform any spatial decomposition of the structure into local and nonlocal subdomains (e.g., PD--FEM coupling across an internal structural interface). Consequently, well-known dynamic artifacts reported for local--nonlocal \emph{structural} coupling, such as spurious wave reflections at a PD--local interface \cite{kulkarni2018study,giannakeas2019wave}, are not expected to arise in the present formulation.

Nonetheless, peridynamic models require careful treatment near physical boundaries due to neighborhood truncation, which can influence localized fields (and, in some cases, the static response) in the vicinity of the boundary. In the present thickness-integrated 1D peridynamic Euler--Bernoulli beam formulation, the beam ends are subject to \emph{kinematic} clamped constraints (displacement and rotation) rather than traction boundary conditions. These constraints are enforced using ghost (fictitious) material points and mirror-image relations for the displacement field near each end, consistent with established peridynamic beam implementations \cite{yang2020,yang2022a}.

For completeness, we note that when traction-type boundary conditions must be imposed in peridynamics (typically in 2D/3D solid formulations), a common approach is the method of layers, in which the prescribed traction is represented by an equivalent body-force density over a fictitious boundary layer \cite{madenci2013peridynamic}. More recent work proposes alternative treatments that improve compatibility with local boundary-condition enforcement \cite{huang2019revisiting}. Such traction-imposition strategies are not required in the present 1D clamped-beam setting but may be relevant for future extensions.

In this work, we model the fluid using classical lubrication theory. This choice is based on the targeted applications of microfluidic flows, in which channels are long and slender, and easily fit into the thin-gap, low-Reynolds-number regime of interest. In these regimes and slender geometries, lubrication theory provides an accurate and computationally efficient reduced-order description of incompressible viscous flow. Our primary motivation for employing peridynamics is on the structural side, where the nonlocal peridynamic beam formulation naturally accommodates damage initiation and discontinuities, which are central to the aims of the present study. Replacing the lubrication equation with a fully peridynamic (or more broadly, nonlocal) fluid solver would substantially increase computational cost and algorithmic complexity (e.g., nonlocal enforcement of incompressibility and boundary-condition treatments), while obscuring the primary focus of this work.

Furthermore, in our simulations, we enforce the assumptions of lubrication theory through the choice of dimensionless parameters and by monitoring the maximum wall slope during time integration; if the slope exceeds a prescribed admissible threshold, the computation is terminated with a warning, indicating that the chosen parameters have entered a regime where the lubrication approximation may no longer be appropriate. Importantly, note that the deformations shown in the figures are dimensionless (scaled), which visually amplifies the apparent magnitude of deflection even when the small-slope condition is satisfied. Yet, perhaps unexpectedly, although classical lubrication is often associated with the small-slope assumptions, it has been shown to remain accurate even when channel shape changes are comparable to the channel height by comparisons to extended lubrication theory and direct numerical simulations \cite{tavakol2017extended}.

In terms of structure, the peridynamic Euler--Bernoulli beam inherits the classical Euler--Bernoulli limitations, including the neglect of transverse shear and kinematic assumptions consistent with small rotations/deflections. The same slope-based stopping criterion is used to prevent interpretation of responses that fall outside this regime within the present model. A natural extension, if larger deformations and shear effects must be captured, is to adopt a shear-deformable beam theory (e.g., a peridynamic Timoshenko beam formulation) \cite{yang2020peridynamic,naumenko2025first}.

Finally, PD and nonlocal-operator formulations for fluid flow and transport phenomena are active areas of research \cite{gao2019nonlocal,mikata2021peridynamics,zhao2022construction}. 
These developments motivate promising future work toward a nonlocal analogue of the Reynolds lubrication equation within the thin-film asymptotic reduction.

\section{Conclusion}
\label{sec:conclusion}

We developed a 1D model, coupling lubrication theory for viscous flow to a nonlocal Euler--Bernoulli beam based on peridynamic theory, to simulate the dynamics, up to failure, of the flow-induced deformation of a channel's compliant top wall. The key novelty of the model is to capture the nonlocal material response of the soft elastic wall using the theory of peridynamics, introducing as a parameter the dimensionless horizon~$\Delta$. Additionally, a Reynolds number $Re$ quantifying flow inertia, a Strouhal number $St$ quantifying unsteady inertia of the beam, and a compliance number $\beta$ quantifying the fluid--structure coupling emerge from the model.

We benchmarked the proposed fluid--peridynamic structure model against previous results from the literature by considering $\Delta \to 0$ at steady-state. Next, dynamic simulations revealed dispersive oscillations and damping in the response of this coupled system. A linearized (dispersion) analysis of the coupled fluid--structure problem revealed that, for any nonzero horizon $\Delta$, the phase velocity of harmonic disturbances departs from the Euler--Bernoulli prediction at dimensionless wavenumbers $k\Delta=\mathcal{O}(1)$ and approaches a finite asymptote at large $k$. This finding demonstrates the intrinsic high-wavenumber filtering produced by the underlying nonlocal elasticity model. 

Beyond wave dynamics, we explored the potential failure of the beam-like soft wall of the microchannel under static and dynamic conditions. We found that a boundary in the $(St,\beta)$ parameter space separates two failure-risk regimes: above this curve, transient loading is more severe than the corresponding steady load, whereas below it, failure is likely under the steady-state load. The location of an ``operating point'' relative to this boundary, therefore, indicates whether the design conditions of microchannel operation are likely to lead to failure during its transient response to the flow or not.

Finally, in this work, we exploited the reduction to a 1D problem via depth-averaged flow. In future work, it would be of interest to solve a 2D problem for the failure of the microchannel wall, including resolving the fracture of the wall and the subsequent dynamics of the escaping fluid. Such a simulation could be performed using the novel immersed-boundary fluid--peridynamic structure interaction solver developed by Kim et al.~\cite{Kim2023}.

\section*{CRediT authorship contribution statement}
\textbf{Ziyu Wang:} Conceptualization, Methodology, Software, Validation, Formal analysis, Investigation, Data Curation, Writing - Original Draft, Writing - review \& editing, Visualization. \textbf{Ivan C.\ Christov:} Conceptualization, Methodology, Investigation, Writing - Original Draft, Writing - review \& editing, Supervision, Project administration, Funding acquisition.

\section*{Declaration of competing interest}
The authors declare that they have no known competing financial interests or personal relationships that could have appeared to influence the work reported in this paper.

\section*{Acknowledgements}
This research was supported by the U.S.\ National Science Foundation under grant CMMI-2245343.

\section*{Data availability}
The data that support the findings of this study are available from the corresponding author upon reasonable request.

\appendix

\setcounter{figure}{0}

\section{Numerical scheme}
\label{subsec:NumericalScheme}
 
In this appendix, we introduce the numerical scheme employed to solve the dynamic coupled problem of peridynamic beam deformation and fluid flow. Our approach is based on the numerical scheme for coupled problems involving classical beam deformation and flow developed in \cite{wang2022reduced}. Let the time‐step size be denoted by $\Delta T$. In this approach, the compliant channel wall is modeled as a PD beam discretized into a single array of uniformly distributed particles, each corresponding to a discrete domain volume. We assign each particle $k$ a volume $\mathcal{V}_k$ (with unit cross‐sectional area), mathematically equal to $\Delta X$. In this formulation, a subscript refers to the function value at the corresponding particle, whereas a superscript indicates the discrete time step index.

The PD beam's equation of motion, Eq.~\eqref{eq:pdbeam_nd}, can be discretized \cite{yang2020} as follows:
\begin{multline}
    St \ddot{H}_{k} 
    - 
    \frac{1}{\Delta^2} 
    \sum_{j} \frac{\mathcal{V}_{j}}{\Xi^2_{j,k}}
    \left(
      \sum_{i} \frac{H_{i^k} - H_{k}}{\Xi^2_{i^k,k}} \mathcal{V}_{i^k} - 
      \sum_{i} \frac{H_{i^j} - H_{j}}{\Xi^2_{i^j,j}} \mathcal{V}_{i^j}
    \right)  \\
    = 
    \beta P_{k},
    \label{eq:pdbeam_discr}
\end{multline}
where $i$ and $j$ are dummy indices, $i^k$ refers to other particles within the horizon of particle $k$, and $\Xi_{j,k}$ denotes the bond length between particles $j$ and $k$. For time step $n+1$, Eq.~\eqref{eq:pdbeam_discr} can be rewritten in matrix form as
\begin{equation}
    St \bm{I} \ddot{H}^{n+1} - 
    \bm{K} H^{n+1} = 
    \beta P^{n+1}.
\end{equation}
Here, $\dot{H}$ represents the vector of vertical velocities of the ``particles'' along the PD beam, with $\dot{(\,\cdot\,)} = \partial(\,\cdot\,)/\partial t$ and $\ddot{(\,\cdot\,)} = \partial^{2}(\,\cdot\,)/\partial t^{2}$. 

In this formulation, the fictitious domain introduced to satisfy the Neumann BC is chosen with a width of $2\Delta$, i.e., twice the peridynamic horizon \cite{yang2022a,madenci2013peridynamic}, so that the displacement field in the adjoining $2\Delta$-wide fictitious region is the mirror image of the interior, thereby enforcing a zero first derivative at the boundary.

The Newmark-$\beta$ scheme \cite{subbaraj1989survey} is used to discretize the beam motion equation \eqref{eq:pdbeam_nd} in time. Specifically, the spatially discretized equation can be solved by sequentially obtaining the acceleration $\ddot{H}$, velocity $\dot{H}$, and the beam height $H$. In matrix form, we have:
\begin{subequations}\begin{empheq}[left=\empheqlbrace]{align}
    \ddot{H}^{n+1}
    &=
    \left(
      St \bm{I} +
      \phi_{2}\Delta T^{2}\bm{K}
    \right)^{-1} \Big\{
      \beta P^{n+1} 
       \\
      &\quad  - 
      \bm{K}
      \left[
        H^{n} + 
        \Delta T \dot{H}^{n} + 
        \Bigl(\tfrac{1}{2} - \phi_{2}\Bigr) \Delta T^{2} \ddot{H}^{n}
      \right]
    \Big\}, \nonumber \\
    \dot{H}^{n+1}
    &=
    \dot{H}^{n} + 
    (1 - \phi_{1})\Delta T\ddot{H}^{n} + 
    \phi_{1}\Delta T \ddot{H}^{n+1},\\
    H^{n+1}
    &=
    H^{n} + 
    \Delta T \dot{H}^{n} + 
    \left(\tfrac{1}{2} - \phi_{2}\right) \Delta T^{2} \ddot{H}^{n} \\
    &\quad +\phi_{2} \Delta T^2 \ddot{H}^{n+1}. \nonumber
\end{empheq}\label{eq:discrete_Hs}\end{subequations}
This scheme is unconditionally stable and exhibits second‐order accuracy when $\phi_{1} = 1/2$ and $\phi_{2} = 1/4$. However, choosing $\phi_{1} > 1/2$ helps suppress numerically induced high-frequency oscillations \cite{subbaraj1989survey}. In our simulations, we used $\phi_{1} = 1.0$ and $\phi_{2} = 0.5625$.

Meanwhile, for the flow equations, we adopt a second‐order backward difference formula in time. The dimensionless lubrication Eqs.~\eqref{eq:continuity_nd} and \eqref{eq:momentum_nd} are discretized and integrated in space (using the trapezoidal rule) \cite{inamdar2020unsteady,wang2022reduced} to yield:
\begin{subequations}\begin{empheq}[left=\empheqlbrace]{align}
    Q_{k}^{n+1} &= 
    1 - 
    \int_{0}^{X_{k}} 
      \dot{H}^{n+1}
      \,d\tilde{X},
\label{eq:Qint}\\
    P_{k}^{n+1} &= 
    \int_{1}^{X_{k}}
    \Bigg[
      -\frac{Re}{H^{n+1}_{k}}
      \left(
        \frac{3Q_{k}^{n+1} - 4Q_{k}^{n} + Q_{k}^{n-1}}{2\Delta T} 
      \right) \nonumber\\
        & \qquad + 
        \frac{6}{5}\frac{\left(\tfrac{Q^{2}}{H}\right)^{n+1}_{k+1} - 
          \left(\tfrac{Q^{2}}{H}\right)^{n+1}_{k-1}}{2\Delta X} - 
      12\left(\frac{Q}{H^{3}}\right)^{n+1}_{k}
    \Bigg]
    d\tilde{X}.
\label{eq:Pint}
\end{empheq}\label{eq:QPint}\end{subequations}
The corresponding BCs~\eqref{eq:flowbc_nd} at the domain edges have been enforced in Eqs.~\eqref{eq:QPint} through the choice of integration limits.

The fully coupled solution procedure advances one time step at a time. Within each time step, we use fixed-point iterations to linearize the nonlinear algebraic problem. First, we initialize the intermediate pressure $P^{I}$ as $P^{n}$ from the previous step and solve for $\ddot{H}^{I}$, $\dot{H}^{I}$, and $H^{I}$ via Eqs.~\eqref{eq:discrete_Hs}. Next, we compute the intermediate flow rate $Q_{k}^{I}$ via Eq.~\eqref{eq:Qint}. Then, we obtain the corrected pressure $P^{n+1}$ via Eq.~\eqref{eq:Pint}. We define a residual $R_d = \| P^{n+1} - P^{I}\|_\infty = \max_k | P^{n+1}_k - P^{I}_k|$, and iterate the latter steps until the relative change of $R_d$ is reduced below a prescribed tolerance of $10^{-5}$. To ensure convergence of the displacement with a relative error of $10^{-6}$ in pure beam bending, the discretization employs over 200 particles along the beam. The time step $\Delta T = 10^{-6}$ is selected to be sufficiently small to guarantee convergence of the simulation results at each time point, also achieving a relative error of $10^{-6}$. The scheme was implemented in \textsc{Matlab}.

\section{Linearization} \label{subsec:Linearization}

In this appendix, we provide details of the linearization of the coupled problem, which enables the dispersion analysis in Sec.~\ref{subsec:dispersion} above. To facilitate a stability analysis, we begin by linearizing the governing equations around a steady base state. In particular, we decompose the primary variables into their base state (zero subscript) and a small perturbation component (primes). Next, we substitute the assumed form of the solution into the governing equations, keeping only linear terms in the primed (perturbation) variables. 

\subsection{Equations for the perturbations}

First, substituting Eqs.~\eqref{eq:HQP_perturb} into the dimensionless peridynamic beam equation~\eqref{eq:pdbeam_nd}, and acknowledging that the base state (zero subscripts) solves the original problem, we obtain
\begin{multline}
   St\frac{\partial^2 H'}{\partial T^2} - \frac{1}{\Delta^2} \int_{-\Delta}^{+\Delta} \frac{1}{\Xi^2} \left[\int_{-\Delta}^{+\Delta}\frac{H'(X+\Theta,T)-H'(X,T)}{\Theta^2}\,d\Theta \right. \\
   \left. - \int_{-\Delta}^{+\Delta}\frac{H'(X+\Xi+\Theta,T)-H'(X+\Xi,T)}{\Theta^2}\,d\Theta \right]\,d\Xi = \beta P',
\label{eq:pdbeam_nd_pert}
\end{multline}
which relates the beam deflection perturbation $H'$ to the pressure perturbation $P'$, with the Strouhal number $St$ and the compliance number $\beta$ scaling the inertial and pressure loading effects, respectively.

Similarly, substituting Eqs.~\eqref{eq:HQP_perturb} into the fluid's continuity equation \eqref{eq:continuity_nd}
and acknowledging that the base state (zero subscripts) solves the original problem, we obtain
\begin{equation}
\frac{\partial Q'}{\partial X} + \frac{\partial H'}{\partial T} = 0,
\label{eq:continuity_nd_pert}
\end{equation}
ensuring mass conservation in the perturbed flow field.

Finally, substituting Eqs.~\eqref{eq:HQP_perturb} into the fluid's momentum equation \eqref{eq:momentum_nd}, we have
\begin{multline}
Re \left\{ \frac{\partial \left(Q_0+Q'\right)}{\partial T} + \frac{6}{5} \frac{\partial}{\partial X} \left[\frac{(Q_0+Q')^2}{H_0+H'}\right]\right\}  \\ 
= -\left(H_0+H'\right) \frac{\partial \left(P_0+P'\right)}{\partial X} - 12\frac{Q_0+Q'}{(H_0+H')^2}.
\label{eq:momentum_nd_exp}
\end{multline}
Unlike the beam and continuity equations, the momentum equation is nonlinear. Thus, we must linearize the nonlinear terms in Eq.~\eqref{eq:momentum_nd_exp} to proceed. We use the following Taylor series approximations:
\begin{subequations}\begin{align}
\frac{(Q_0+Q')^2}{H_0+H'} &\approx \frac{Q_0^2}{H_0} + \frac{2Q_0Q'}{H_0} - \frac{Q_0^2H'}{H_0^2}, \\
\frac{Q_0+Q'}{(H_0+H')^2} &\approx \frac{Q_0}{H_0^2} - 2\frac{Q_0H'}{H_0^3} + \frac{Q'}{H_0^2}.
\end{align}\label{eq:Taylor_series}\end{subequations}
Then, using the approximations from Eqs.~\eqref{eq:Taylor_series} and subtracting Eq.~\eqref{eq:momentum_nd} from Eq.~\eqref{eq:momentum_nd_exp}, the momentum equation for the perturbation is
\begin{multline}
Re \left[ \frac{\partial Q'}{\partial T} + \frac{6}{5} \frac{\partial}{\partial X} \left( \frac{2Q_0Q'}{H_0} - \frac{Q_0^2 H'}{H_0^2} \right) \right] \\
= -H'\frac{\partial P_0}{\partial X} - H_0\frac{\partial P'}{\partial X} - 12\left(-2\frac{Q_0 H'}{H_0^3} + \frac{Q'}{H_0^2}\right).
\label{eq:momentum_nd_pert}
\end{multline}

\subsection{Transformation into a linear system}
\label{subsec:Transformation}

We next assume harmonic solutions for the perturbations, as in Eqs.~\eqref{eq:HQP_perturb}, which allows the partial derivatives with respect to $X$ and $T$ to be replaced by the algebraic factors $ik$ and $-i\omega$, respectively.

Substituting these harmonic solutions into the linearized peridynamic beam equation~\eqref{eq:pdbeam_nd_pert} yields
\begin{multline}\label{eq:pdbeam_nd_pert_2}
   -St\omega^2 \hat{H}e^{i(kX-\omega T)} \\
   - \frac{1}{\Delta^2}\int_{-\Delta}^{+\Delta} \frac{1}{\Xi^2} \left[\int_{-\Delta}^{+\Delta}\frac{\hat{H}e^{i(k(X+\Theta)-\omega T)}-\hat{H}e^{i(kX-\omega T)}}{\Theta^2} \,d\Theta \right. \\ \left. - \int_{-\Delta}^{+\Delta}\frac{\hat{H}e^{i(k(X+\Xi+\Theta)-\omega T)}-\hat{H}e^{i(k(X+\Xi)-\omega T)}}{\Theta^2} \,d\Theta \right] \,d\Xi \\
   = \beta \hat{P}e^{i(kX-\omega T)}.
\end{multline}
Then we perform the necessary algebraic manipulations and evaluate the nonlocal integral terms:
\begin{multline}
   \int_{-\Delta}^{+\Delta}\frac{\hat{H}e^{i(k(X+\Theta)-\omega T)}-\hat{H}e^{i(kX-\omega T)}}{\Theta^2} \,d\Theta \\
   = \hat{H}e^{i(kX-\omega T)}\int_{-\Delta}^{+\Delta}\frac{e^{ik\Theta}-1}{\Theta^2} \,d\Theta \\ 
   = \hat{H}e^{i(kX-\omega T)} \left( \int_{-\Delta}^{+\Delta}\frac{\cos(k\Theta)+i\sin(k\Theta)}{\Theta^2} \,d\Theta - \int_{-\Delta}^{+\Delta}\frac{1}{\Theta^2} \,d\Theta \right) \\ =  
   \hat{H}e^{i(kX-\omega T)} \left( \int_{-\Delta}^{+\Delta}\frac{\cos(k\Theta)}{\Theta^2} \,d\Theta + \frac{2}{\Delta}\right)
\end{multline}
and, similarly,
\begin{multline}
  \int_{-\Delta}^{+\Delta}\frac{\hat{H}e^{i(k(X+\Xi+\Theta)-\omega T)}-\hat{H}e^{i(k(X+\Xi)-\omega T)}}{\Theta^2} \,d\Theta \\ = \hat{H}e^{i(kX+k\Xi-\omega T)} \left( \int_{-\Delta}^{+\Delta}\frac{\cos(k\Theta)}{\Theta^2} \,d\Theta + \frac{2}{\Delta}\right).
\end{multline}
Thus,
\begin{multline}\label{eq:double_ints}
\int_{-\Delta}^{+\Delta} \frac{1}{\Xi^2} \left[\int_{-\Delta}^{+\Delta}\frac{\hat{H}e^{i(k(X+\Theta)-\omega T)}-\hat{H}e^{i(kX-\omega T)}}{\Theta^2} \,d\Theta \right. \\ \left.
- \int_{-\Delta}^{+\Delta}\frac{\hat{H}e^{i(k(X+\Xi+\Theta)-\omega T)}-\hat{H}e^{i(k(X+\Xi)-\omega T)}}{\Theta^2} \,d\Theta \right] \,d\Xi \\
= 
-\left( \int_{-\Delta}^{+\Delta}\frac{\cos(k\Theta)}{\Theta^2} \,d\Theta + \frac{2}{\Delta}\right) \\
\times \int_{-\Delta}^{+\Delta} \frac{\hat{H}e^{i(kX+k\Xi-\omega T)}-\hat{H}e^{i(kX-\omega T)} }{\Xi^2}  \,d\Xi \\ 
= 
-\hat{H}e^{i(kX-\omega T)}\left( \int_{-\Delta}^{+\Delta}\frac{\cos(k\Theta)}{\Theta^2} \,d\Theta + \frac{2}{\Delta}\right)^2.
\end{multline}
Now, note that
\begin{equation}\label{eq:cos_int}
\begin{aligned}
 &\int_{-\Delta}^{+\Delta}\frac{\cos(k\Theta)}{\Theta^2} \,d\Theta 
 \\
 &= \int_{-\Delta}^{+\Delta}\frac{1}{\Theta^2} \left(1-\frac{k^2\Theta^2}{2!}+\frac{k^4\Theta^4}{4!}- \cdots\right) \,d\Theta \\ 
 &=  
 \int_{-\Delta}^{+\Delta}\frac{1}{\Theta^2} -\frac{k^2}{2!}+\frac{k^4\Theta^2}{4!}- \frac{k^6\Theta^4}{6!} + \cdots \,d\Theta \\ 
 &= 
 -\frac{2}{\Delta} + \left. \left(-\frac{k^2\Theta} {2!}+\frac{k^4\Theta^3}{3\cdot4!}- \frac{k^6\Theta^5}{5\cdot6!} + \cdots\right) \right\rvert_{-\Delta}^{+\Delta} \\ 
 &= 
 -\frac{2}{\Delta} + 2k\sum_{n=1}^{\infty} \frac{(-1)^n(k\Delta)^{2n-1}}{(2n-1)(2n)!}.
\end{aligned}
\end{equation}
Using Eq.~\eqref{eq:cos_int}, Eq.~\eqref{eq:double_ints} becomes
\begin{multline}
\int_{-\Delta}^{+\Delta} \frac{1}{\Xi^2} \left[\int_{-\Delta}^{+\Delta}\frac{\hat{H}e^{i(k(X+\Theta)-\omega T)}-\hat{H}e^{i(kX-\omega T)}}{\Theta^2} d\Theta \right. \\ \left. - \int_{-\Delta}^{+\Delta}\frac{\hat{H}e^{i(k(X+\Xi+\Theta)-\omega T)}-\hat{H}e^{i(k(X+\Xi)-\omega T)}}{\Theta^2} d\Theta \right] d\Xi \\ =  
-\hat{H}e^{i(kX-\omega T)}\left( 2k\sum_{n=1}^{\infty} \frac{(-1)^n(k\Delta)^{2n-1}}{(2n-1)(2n)!} \right)^2,
\end{multline}
which can be substituted into Eq.~\eqref{eq:pdbeam_nd_pert_2} to obtain
\begin{multline}
   -St\omega^2 \hat{H}e^{i(kX-\omega T)} \\
   + \frac{1}{\Delta^2} \hat{H}e^{i(kX-\omega T)}\left( 2k\sum_{n=1}^{\infty} \frac{(-1)^n(k\Delta)^{2n-1}}{(2n-1)(2n)!} \right)^2 \\
   = \beta \hat{P}e^{i(kX-\omega T)}.
\end{multline}
The latter equation easily simplifies to the final form of the linearized peridynamic beam equation:
\begin{equation}
\left[-St\,\omega^2 + \frac{1}{\Delta^2} \left( 2k\sum_{n=1}^{\infty}\frac{(-1)^n (k\Delta)^{2n-1}}{(2n-1)(2n)!} \right)^2 \right]\hat{H} - \beta \hat{P} = 0,
\end{equation}
which yields  Eq.~\eqref{eq:momentum_nd_pert_harmonic} above.

Similarly, substituting the harmonic form of the perturbations into the linearized continuity equation~\eqref{eq:continuity_nd_pert} immediately yields
\begin{equation}
ik\hat{Q} - i\omega\,\hat{H} = 0,
\end{equation}
which yields  Eq.~\eqref{eq:continuity_nd_pert_harmonic} above.

Finally, we substitute the harmonic form of the perturbation into the linearized momentum equation~\eqref{eq:momentum_nd_pert}, to obtain
\begin{multline}
Re \left[ -i\omega \hat{Q}e^{i(kX-\omega T)} \right. \\
\left. + \frac{6}{5} ik \left( \frac{2Q_0}{H_0}\hat{Q}e^{i(kX-\omega T)} - \frac{Q_0^2}{H^2_0}\hat{H}e^{i(kX-\omega T)} \right) \right] \\ 
= 
-ik H_0 \hat{P}e^{i(kX-\omega T)} - 12 \left(- 2\frac{Q_0\hat{H}e^{i(kX-\omega T)}}{H^3_0} +\frac{\hat{Q}e^{i(kX-\omega T)}}{H^2_0} \right).
\end{multline}
By collecting like terms, the last equation can be rearranged into the final algebraic form:
\begin{multline}
\left[Re \left( -i\omega + \frac{12}{5}\frac{ikQ_0}{H_0} \right) + \frac{12}{H_0^2}\right]\hat{Q} \\
+ \left(-ik Re \frac{6}{5}\frac{Q_0^2}{H_0^2} - 24\frac{Q_0}{H_0^3}\right)\hat{H} 
+ ikH_0\hat{P} = 0,
\end{multline}
as given in Eq.~\eqref{eq:momentum_nd_pert_harmonic} above.

\balance

\setlength{\bibsep}{1pt}
{\footnotesize
\bibliographystyle{elsarticle-num-names} 
\bibliography{references.bib}
}

\end{document}